
%
%

\input jnl-0.3.tex


\font\titlefont=cmr10 scaled \magstep3
\def\bigtitle{\null\vskip 3pt plus 0.2fill \beginlinemode \doublespace
\raggedcenter \titlefont}

\gdef\journal#1, #2, #3, 1#4#5#6{
{\sl #1~}{\bf #2}, #3 (1#4#5#6)}

\def\a{\rightarrow}
\def\aa{\alpha}
\def\b{\beta}
\def\c{\cosh \alpha}

\def\g{\gamma}

\def\p{\partial}
\def\r{\hat r}
\def\s{\sinh \alpha}
\def\slr{SL(2,\R)}
\def\ss{\sinh^2 \alpha}
\def\t{\tilde}

\def\P{P_x}
\def\R{{\bf R}}

\def\+{{\scriptscriptstyle +}}
\def\-{{\scriptscriptstyle -}}
\singlespace
\preprintno{UCSBTH-91-53}
\preprintno{October, 1991}
\doublespace
\bigtitle An Equivalence Between Momentum and Charge in String Theory

\author James H. Horne, Gary T. Horowitz, and Alan R. Steif

\affil
        Institute for Theoretical Physics
        University of California
        Santa Barbara, CA 93106-4030
\centerline{jhh@cosmic.physics.ucsb.edu}
\centerline{gary@cosmic.physics.ucsb.edu}
\centerline{steif@cosmic.physics.ucsb.edu}

\abstract
It is shown that for a translationally invariant solution to string
theory, spacetime duality interchanges the momentum in the symmetry
direction and the axion charge per unit length.  As one application,
we show explicitly that charged black strings are equivalent to
boosted (uncharged) black strings. The extremal black strings (which
correspond to the field outside of a fundamental macroscopic string)
are equivalent to plane fronted waves describing strings moving at the
speed of light.

\endtitlepage
\baselineskip=16pt

Despite extensive work over the past several years, we still lack a
fundamental description of string theory, including a complete
understanding of the basic objects that are involved and the
principles which guide its construction. However there are several
hints and suggestive clues which have been uncovered. Perhaps the most
important is spacetime
duality~[\cite{Kikkawa},\cite{Buscher},\cite{Smith}].  This is the
fact that different spacetime backgrounds correspond to equivalent
solutions in string theory. Spacetime duality is usually thought of in
terms of relating small distances to large distances since in the
simplest example of flat space with one direction identified, the
duality relates radius $R$ to radius $\alpha'/R$.\footnote*{In this
paper, we will set $\alpha' = 1$. (In ref.~[\cite{Horne}],
$\alpha'={1\over 2}$ was used.)} Although this is certainly an
important aspect of duality, there are many other consequences as
well. For example, it has been shown that spacetime manifolds of
different topology can be equivalent as string solutions (for a recent
review, see~[\cite{Greene}]).  In this paper we describe another
consequence of duality which concerns asymptotically defined conserved
quantities.  For a solution which has a translational symmetry and is
asymptotically flat in the transverse directions, one can define the
total energy momentum per unit length $P_\mu$ and the total charge per
unit length $Q$ associated with the three form $H_{\mu\nu\rho}$. (We
will refer to $Q$ as the ``axion charge".) We will show that duality
interchanges $Q$ and the component of $P_\mu$ in the symmetry
direction, while leaving the orthogonal components of $P_\mu$
unchanged. Thus, in this sense, linear momentum is equivalent to axion
charge in string theory.

An immediate consequence of this equivalence is that one can add axion
charge to any solution which is static as well as translationally
invariant as follows: First boost the solution to obtain a nonzero
linear momentum and then apply a duality transformation to convert
this momentum to charge.  (This procedure is related to the
transformations discussed in ref.~[\cite{Sen}].)  We will illustrate
this by considering the black string solutions.  Black strings are one
dimensional extended objects surrounded by event horizons.  Solutions
to the low energy field equations describing charged black strings in
ten dimensions were found in ref.~[\cite{Horowitz}].  Exact conformal
field theories describing charged black strings in three dimensions
were found in ref.~[\cite{Horne}]. We will consider these three
dimensional solutions first and describe the duality between charge
and momentum.  For $|Q|<M$, duality relates solutions with an inner
horizon and a timelike singularity to solutions with no inner horizon
and a spacelike singularity. For $|Q|>M$, duality relates completely
nonsingular solutions to solutions with naked singularities.  We then
construct charged black string solutions in higher dimensions $D \ge
5$ by boosting the uncharged solution (which is simply the product of
a $D-1$ dimensional black hole and the real line) and applying the
duality transformation.

It is known~[\cite{Horowitz}] that the field outside of a straight
fundamental string in ten dimensions~[\cite{Dabholkar}] is just the
extremal black string ($|Q|=M$). We will see that this is true in all
dimensions $D\ge 5 $. More importantly, the extremal black strings
turn out to be equivalent under duality to plane fronted waves, which
describe strings moving at the speed of light.  We conclude with some
further consequences of the linear momentum -- axion charge
equivalence.

We begin by reviewing the definition of energy-momentum and axion
charge. Since they are defined as surface integrals at infinity where
the fields are weak, it suffices to work with the low energy equations
of motion.  These can be obtained by varying the action
$$ S = \int e^\phi[R + (\nabla\phi)^2 - {1\over 12}H^2 + a^2].
                                                    \eqno(lowenergy)$$
where $H=dB$ is the three form and
$a^2$ denotes a possible cosmological constant.
Let
$g_{\mu\nu}$, $H_{\mu\nu\rho}$ and $\phi$ be an extremum of $S$
(in $D$ dimensions)
which is independent of a coordinate
$x$. We assume that at large distances transverse to $x$,
$g_{\mu\nu} \a \eta_{\mu\nu} + \gamma_{\mu\nu}$,
$B_{\mu\nu} \a 0$ and $\phi \a \phi_0 + \chi$.
The field equations require that $\phi_0 = ay + b$ for some Cartesian
coordinate $y$.
There is a conserved component of the energy momentum associated with
each translational symmetry of the background. (Thus, if $a \ne 0$,
$P_y$ is not well defined.) For $D\ge 5$, we require
$a=0$ and $\g_{\mu\nu}, B_{\mu\nu}, \chi$ to all fall off like $r^{4-D}$
where $r$ is a radial coordinate orthogonal to $x$. For $D=3$, we allow
$a \ne 0 $ and require $\g_{\mu\nu}, B_{\mu\nu}, \chi$ to fall off like
$e^{-ar}$.
The total energy momentum is obtained by first extremizing
eq.~\(lowenergy) with respect to the metric and linearizing the
resulting expression about the background solution.  One then
contracts with a translational symmetry and integrates over a
spacelike plane. The result is a total derivative which can be
reexpressed as a surface integral at infinity. Due to the
translational symmetry in $x$, it is convenient to work with the
energy momentum per unit length.  Its time component (the ADM energy
per unit length) is given by\footnote*{This formula is valid in an
asymptotically Cartesian coordinate system. To obtain a formula valid
in any coordinate system, one simply replaces the partial derivatives
with covariant derivatives.}
$$ M= C\oint e^{\phi_0} [ \p^i\g_{ij} - \p_j(\g + 2\chi)
 + \g_{ij} \p^i \phi_0] dS^j \eqno(massdef)$$
where $C$ is a dimension dependent constant,
$i,j$ run over spatial indices, $\g = \g^i_i$, and the integral is over the
$D-3$ sphere at large $r$ and constant $x$.
The spatial components are
$$  P_i = C\oint e^{\phi_0} [\p_t \g_{ij} - \p_j \g_{ti} + \p_i \g_{tj} ] dS^j
                        . \eqno(momidef) $$
Since $\p_x \g_{\mu\nu} = 0 $, we have
$$  P_x = C\oint e^{\phi_0} [\p_t \g_{xr} - \p_r \g_{tx} ] .\eqno(momxdef)$$
Finally the total charge per unit length associated with
$H$ is given by
$$     Q=C\oint e^{\phi_0} *H = C\oint e^{\phi_0} H_{xrt}
        =C\oint e^{\phi_0} [\p_t B_{xr} + \p_r B_{tx}]  \eqno(qdef)$$
where we have chosen a gauge with $\p_x B_{rt} = 0$ in the last step.

Given a solution $g_{\mu\nu}$, $H_{\mu\nu\rho}$, $\phi$ with a
translational symmetry in $x$, there is a dual
solution~[\cite{Buscher}] given by:
$$\eqalign{
 \tilde g_{xx} & = 1/g_{xx}, \qquad  \t g_{x\aa} = B_{x\aa}/ g_{xx} \cr
 \t g_{\aa\b} & = g_{\aa\b} - (g_{x\aa}g_{x\b} - B_{x\aa}B_{x\b})/g_{xx} \cr
 \t B_{x\aa} & = g_{x\aa}/g_{xx}, \qquad
                      \t B_{\aa\b} = B_{\aa\b} -2 g_{x[\aa} B_{\b]x}/g_{xx} \cr
  \tilde \phi & = \phi + \log g_{xx} \cr }\eqno(sigmadual)  $$
where $\aa,\b$ run over all directions except $x$.  This is sometimes
referred to as ``sigma model duality".  This map between solutions of
the low energy field equations exists whether or not $x$ is compact.
It has recently been shown~[\cite{Rocek}] that if $x$ is compact, the
original solution and its dual are both low energy approximations to
the {\it same} conformal field theory\footnote*{If $x$ is not compact,
there is an equivalent dual description based on eq.~\(sigmadual), but
one must view $x$ in the dual solution as having zero radius so there
are no momentum modes and only winding
modes~[\cite{Rocek},\cite{Polchinski}].}.  With the above asymptotic
conditions we see that under this duality transformation, the
asymptotic fields transform as
$$\eqalign{ \t \g_{xx} = -\g_{xx} &\qquad \t \chi = \chi + \g_{xx} \cr
 \t \g_{x\aa} = B_{x\aa}  &\qquad \t B_{x\aa} = \g_{x\aa} \>. }\eqno(asympt)$$
The remaining fields are unchanged to leading order. It is now easy
to verify that under duality, $\P$ and $Q$ are interchanged while the
orthogonal components of $P_\mu$ are unchanged.

We now use this duality to generate the three dimensional charged
black strings which were constructed in ref.~[\cite{Horne}].  Since
the underlying conformal field theory is known, we will indicate how
the duality between the charged solutions and the boosted uncharged
solutions is represented in the exact solution.  First, let us
construct the $|Q|<M$ black string. We begin with the product of the
two dimensional Lorentzian black hole~[\cite{Witten}] with $\R$,
corresponding to the uncharged black string in three dimensions.  The
low energy metric is
$$\eqalign{ ds^2 = -\left(1 - {M_0 \over r} \right) d\hat t^2 +
           & { k \over 4 r^2 (1 - M_0/r)} dr^2 + d\hat x^2 \cr
\phi = \log r + {1 \over 2} \log k, \quad & \quad B = 0 , \cr} \eqno(unthree)$$
where $k$ is related to the cosmological constant $a$.
(A simple coordinate
transformation $r = M_0 \cosh^2 \r$ puts eq.~\(unthree) in
the more familiar form
$k\, d\r^2 - \tanh^2 \r \,d\hat t^2 + d\hat x^2$.)
We now apply a Lorentz boost
$\hat t = t \cosh \alpha + x \sinh \alpha$ and
$\hat x = x \cosh \alpha + t \sinh \alpha$
to the solution~\(unthree)
to obtain a new solution
$$\eqalign{ ds^2 = &-\left(1 - {M_0 \cosh^2 \alpha\over r} \right) dt^2 +
            { k \over 4 r^2 (1 - M_0 / r)} dr^2 \cr &
         +  {2 M_0 \cosh \alpha \sinh \alpha\over r}  dt dx +
           \left(1 + {M_0 \sinh^2 \alpha \over r}\right) dx^2 \cr
\phi = & \log r + {1 \over 2} \log k, \quad  \quad B = 0} \eqno(unboost)$$
which has energy per unit length $M=M_0\cosh^2 \alpha$  and momentum
per unit length $P_x = M_0 \cosh \alpha \sinh \alpha$.\footnote*{Note
that even though the two solutions are
related by a coordinate transformation, they are physically different
since the coordinate transformation does not reduce to the identity
at infinity.} We now apply the duality transformation~\(sigmadual)
on $x$ to obtain
(after shifting $r \a r - M_0 \sinh^2 \alpha$)
$$\eqalign{  ds^2 = - \left(1-{M\over r}\right) dt^2
	   + & \left(1-{Q^2\over M r}\right)dx^2
	   + \left(1-{M\over r}\right)^{-1} \left(1-{Q^2\over Mr}\right)^{-1}
			{k \, dr^2 \over 4 r^2} \cr
\phi = \log r + {1\over 2} \log k \>, \quad & \quad
B_{xt} = Q/r }\eqno(blackstring)$$
where $M = M_0 \cosh^2 \alpha$ and $Q = M_0 \cosh\alpha \sinh\alpha$.
This is the $|Q|<M$ charged black string in three dimensions~[\cite{Horne}],
and the momentum and charge have been switched under duality.

If $x$ is not periodically identified, then eqs.~\(unboost)
and~\(blackstring) presumably represent different conformal field
theories.  However, if $x$ is compact, these solutions are equivalent.
At first sight this is quite surprising since the global structure of
the two spacetimes is quite different.  As discussed in
ref.~[\cite{Horne}], the charged black string~\(blackstring) has an
inner horizon and a timelike singularity, while eq.~\(unboost) has the
usual event horizon and a spacelike singularity.  However, with
hindsight, this result could have been anticipated.  Inner horizons in
general relativity are known to be
unstable~[\cite{Penrose},\cite{Chandra}].  Since the arguments are
quite general, one expects that they should apply in string theory as
well. (One can show {\it e.g.}~that the tachyon field becomes singular
near the inner horizon in the metric~\(blackstring).)  Now a solution
to the low energy field equations is a good approximation to an exact
solution to string theory only if the fields are weak so the higher
order corrections are initially small, {\it and} if it is stable so
that the corrections remain small. If the solution is unstable the
higher order corrections will become large and change the character of
the solution. Thus even though the low energy metric~\(blackstring)
has an inner horizon, the fact it is equivalent to eq.~\(unboost)
shows the exact solution does not.

The derivation above resulted in $|Q| < M$.  As discussed in
ref.~[\cite{Horne}], the metric~\(blackstring) is nonsingular for $|Q| > M$
as long as $r$ is redefined using $r = \t r^2 + Q^2/M$, and $x$
is identified with the appropriate period.  We can obtain it by
starting with the product of a negative mass Euclidean black hole with
time
$$\eqalign{ ds^2 = & - d\hat t^2 + \left(1 + {M_0 \over r}\right) d\hat\theta^2
      + { k \over 4 r^2 (1 + M_0/ r) } dr^2, \cr
\phi = & \log r + {1 \over 2} \log k, \quad \quad B = 0,} \eqno(uneucl)$$
where $\hat\theta$ is not periodic.
This metric  has a naked singularity at $r = 0$.
Now boost the solution~\(uneucl) using
$\hat t = t \cosh\alpha + \theta \sinh \alpha$,
and $\hat \theta = \theta \cosh\alpha + t\sinh\alpha$ to obtain
$$\eqalign{ds^2 = &- \left( 1 - {M_0 \sinh^2\alpha \over r} \right) dt^2
         + \left( 1 + {M_0 \cosh^2\alpha \over r} \right) d\theta^2 \cr
         &+ {2 M_0 \cosh\alpha\sinh\alpha \over r} dt\,d\theta
     + { k \over 4 r^2 (1 + M_0 / r)} dr^2 \>,\cr
\phi = &\log r + {1 \over 2} \log k, \quad  \quad B = 0.} \eqno(euboost)$$
A duality transformation on $\theta$ in eq.~\(euboost)
along with a shift $r \a r - M_0 \cosh^2\alpha$ again yields the
charged black string~\(blackstring) but now
with $M = M_0 \sinh^2 \alpha$ and $Q= M_0 \cosh\alpha \sinh\alpha$, so that
$|Q| >M$.

Again we see that the global structure of the solutions~\(euboost)
and~\(blackstring) is different.  The boosted solution~\(euboost) is
singular at $r=0$, but the charged solution~\(blackstring) (with
$|Q|>M$) has bounded curvature everywhere.  (Euclidean examples of
this phenomenon have been noticed
previously~[\cite{Dijkgraaf},\cite{Giveon},\cite{Kiritsis},\cite
{Martinec},\cite{Rocek}].) If $\theta$ is periodic, these solutions
are equivalent. This shows that certain curvature singularities in
the low energy metric do not adversely affect string propagation.
However, it is known that there are other types of curvature
singularities which do affect strings very strongly~[\cite{Steif}].
It remains an outstanding question to understand the basic difference
between these types of singularities.

The exact conformal field theories corresponding to these three
dimensional black strings are known in terms of gauged WZW models. In
this case, the duality~\(sigmadual) corresponds to interchanging
vector and axial vector gauging. As shown in ref.~[\cite{Horne}], the
charged black strings are obtained by axial gauging an appropriate one
dimensional subgroup of $\slr\times\R$.  One can verify that the
boosted uncharged black strings are obtained by vector gauging the
same subgroup\footnote*{We have been told that Ginsparg and Quevedo
have made a similar observation.}.

The extremal black string ($|Q|\a M$) does not correspond directly to a WZW
model, but can be realized as a limit of either the $|Q|>M$ or
$|Q|<M$ constructions. This limit can be taken in a variety
of inequivalent ways~[\cite{Horne}]. The limit that preserves
the asymptotic behavior has the low energy solution
$$\eqalign{ ds^2  =  \left(1-{M\over r}\right) (-dt^2  +  dx^2  )
      & +  \left(1-{M\over r} \right) ^{-2}  {k \,dr^2\over 4r^2} \cr
\phi  =  \log r + {1 \over 2} \log k\, , \qquad &
B_{xt}  =  {M\over r} \>. }\eqno(exthree)$$
The correct extension past $r=M$ is given in terms of
a new radial coordinate $\t r^2 = r - M$.
The resulting metric has a horizon but no singularity.
We can dualize on $x$
and then make the coordinate transformation
$$x = {1 \over 2} (u-v) ,\quad
t = {1 \over 2} (u+v) , \quad
r  =  M + e^{ 2 \hat r/\sqrt{k}} , \eqno(exshift)$$
to obtain the following solution
$$\eqalign{ ds^2  = & - du\,dv  \,+\,
  d\hat r^2      +
   M e^{ -2 \hat r /\sqrt{k}}  du^2   \cr
\phi  =  & {2 \over \sqrt{k}} \hat r + {1 \over 2} \log k \, , \qquad
B=0 \>.}
                             \eqno(plane)$$
This metric describes a plane fronted wave.
It corresponds to boosting either
eq.~\(unthree) or eq.~\(uneucl) to the speed of light.
It is possible to show
that eq.~\(plane) is a solution to the string equations of motion to
all orders in
$\alpha^\prime$~[\cite{Guven},\cite{Amati},\cite{Steif}]. This is a
result of the fact that the curvature is null, so that all higher
order curvature terms vanish. (The solution we have obtained is
actually a slight generalization of the plane fronted waves that have
been discussed in the literature, since it includes a linear dilaton.)

Most of our results about black strings in $D=3$ can be extended to $D
\ge 5$\footnote*{There do not appear to be static, asymptotically flat
black strings in four dimensions~[\cite{Horowitz},\cite{Dabholkar}].}.
Since the exact conformal field theories are not known explicitly, we
work with the low energy field equations. These should be a good
approximation to the exact solution away from the singularity.  The
uncharged black string solution for $D\ge 5$ is simply the product of
the higher dimensional Schwarzschild solution with $\R$:
$$\eqalign{
   ds^2 = -(1-M_0/r^n)d\hat t^2 + &(1-M_0/r^n)^{-1}dr^2 +r^2 d\Omega^2_{n+1}
      + d \hat x^2
	     \cr
 \phi = 0, &\qquad B =0 } \eqno(unall) $$
where $n=D-4$.  As before, we boost and dualize. The result is
$$\eqalign{
 ds^2 = -{(1-M_0/r^n)\over (1+M_0\ss/r^n)}dt^2 +&  {dx^2\over (1+M_0\ss/r^n)}
    +{dr^2\over (1-M_0/r^n)} + r^2 d\Omega^2_{n+1} \cr
    \phi =& \log (1+M_0 \ss/r^n)\>,
 \cr
    B_{xt} =& { M_0 \c\s  \over r^n + M_0\ss} \>
. } \eqno(gencharge)$$
This solution describes a charged black string. For $D=10$ ($n=6$)
it is precisely the charged black string
solution found in ref.~[\cite{Horowitz}]. It
has mass $M= M_0(1 + {n\over n+1} \sinh^2 \alpha)$ and charge
$ Q = {n  \over n+1} M_0\c\s$.

The field outside of a fundamental macroscopic string in $D\ge 5$ was
found in ref.~[\cite{Dabholkar}].  It was shown in
ref.~[\cite{Horowitz}] that for $D=10$, this field is simply the
extremal limit $Q=M$ of the black string.  From the above formulas for
$M$ and $Q$, the extremal limit corresponds to taking
$$M_0 \a 0, \qquad  \alpha \a \infty \eqno(exlimit)$$
such that $M_0\ss $ stays constant.  It is easy to verify that for any
$D\ge 5 $, the extremal limit of eq.~\(gencharge) is the field outside
of a fundamental string.  It follows from eq.~\(exlimit) that the
solutions dual to the extremal strings are precisely the black
strings~\(unall) boosted to the speed of light, which are described by
$$  ds^2= -(1-M/r^n)dt^2 + {2 M\over r^n} dt \,dx +(1+  M/r^n)dx^2
 + dr^2 +r^2 d\Omega^2_{n+1}\>. \eqno(exgena)$$
With new coordinates $x={1 \over 2} (u-v)$ and
$ t = {1\over 2}(u+v)$, the metric becomes
$$   ds^2 = - du \,dv + dr^2+r^2 d\Omega^2_{n+1} + {M \over r^n} du^2.
                \eqno(exgenb)$$
This is precisely the form of a plane fronted
wave~[\cite{Guven},\cite{Amati},\cite{Steif}].  If
we were to add a $\delta(u)$ factor to $g_{uu}$, then eq.~\(exgenb)
would be the Aichelburg--Sexl metric describing a point particle
boosted to the speed of light~[\cite{Aichelburg}].  Without this
factor, the metric describes a string boosted to the speed of light.

The equivalence between charge and linear momentum undoubtedly has
broad implications. We have discussed applications to black string
solutions.  We now conclude with two possible further consequences.
First, from the conventional viewpoint, linear momentum is associated
with a spacetime symmetry and axion charge is associated with an
internal symmetry. The fact that they are equivalent is a concrete
indication of a unification of these symmetries in string theory.
Second, if $x$ is compact, $\P$ should be quantized.  This implies
that axion charge must be quantized as well.  This appears to be a new
argument for quantization of charge.

\subhead{Acknowledgements}
We would like to thank
N.~Ishibashi,
M.~Li,
J.~Polchinski, and
A.~Strominger
for
useful discussions.
This work was supported in part by NSF Grant PHY-9008502.

\vfill\eject

\references

\baselineskip=16pt

\refis{Witten} E.~Witten, ``On String Theory and Black Holes,''
\journal Phys. Rev., D44, 314, 1991.

\refis{Dijkgraaf} R.~Dijkgraaf, E.~Verlinde and H.~Verlinde,
``String Propagation in a Black Hole Geometry,'' Princeton
preprint, PUPT-1252, May 1991.

\refis{Martinec} E.~Martinec and S.~Shatasvili, ``Black Hole Physics
and Liouville Theory,'' Enrico Fermi preprint, EFI-91-22, May 1991.

\refis{Kiritsis} E.~Kiritsis, ``Duality in Gauged WZW Models,''
Berkeley preprint, LBL-30747, May 1991.

\refis{Giveon} A.~Giveon, ``Target Space Duality and Stringy Black Holes,''
Berkeley preprint, LBL-30671, April 1991.

\refis{Horowitz} G.~Horowitz and A.~Strominger, ``Black Strings
and $p$-Branes,''
\journal Nucl.~Phys., B360, 197, 1991.

\refis{Chandra} S.~Chandrasekhar and J.~Hartle,
\journal Proc. Roy. Soc. Lond., A384, 301, 1982.

\refis{Steif} G.~Horowitz and A.~Steif,
\journal Phys.~Rev.~Lett., 64, 260, 1990;
\journal Phys.~Rev., D42, 1950, 1990.

\refis{Dabholkar} A.~Dabholkar, G.~Gibbons, J.~Harvey and F.~Ruiz,
``Superstrings and Solitons,''
\journal Nucl. Phys., B340, 33, 1990.

\refis{Horne} J.~Horne and G.~Horowitz, ``Exact Black Strings Solutions
in Three Dimensions,'' UCSB preprint UCSBTH-91-39, July 1991,
to appear in {\it Nucl. Phys.}

\refis{Polchinski} J.~Polchinski, private communication.

\refis{Smith} E.~Smith and J.~Polchinski,
``Duality Survives Time Dependence,'' \journal Phys. Lett., D263, 59, 1991;
A.~Tseytlin, ``Duality and Dilaton,''
\journal Mod. Phys. Lett., A6, 1721, 1991.

\refis{Buscher} T.~Buscher, ``Path Integral Derivation of Quantum
Duality in Nonlinear Sigma Models,''
\journal Phys. Lett., B201, 466, 1988;
``A Symmetry of the String Background Field Equations,''
\journal Phys. Lett., B194, 59, 1987.

\refis{Guven} R.~Guven, \journal Phys. Lett., B191, 275, 1987.

\refis{Amati} D.~Amati and C.~Klimcik, \journal Phys. Lett., B219, 443, 1989.

\refis{Kikkawa} K.~Kikkawa and M.~Yamasaki,
\journal Phys. Lett., B149, 357, 1984;
N.~Sakai and I.~Senda,
\journal Prog. Theor. Phys., 75, 692, 1986;
V.~Nair, A.~Shapere, A.~Strominger and F.~Wilczek,
\journal Nucl. Phys., B287, 402, 1987.

\refis{Greene} B.~Greene and M.~Plesser, ``Mirror Manifolds:
A Brief Review and Progress Report,'' Cornell preprint CLNS 91-1109,
September 1991.

\refis{Penrose} M.~Simpson and R.~Penrose,
\journal Int. J. Theor. Phys., 7, 183, 1973.

\refis{Rocek} M.~Ro\v cek and E.~Verlinde,
``Duality, Quotients, and Currents,'' IAS preprint, IASSNS-HEP-91/68,
October 1991.

\refis{Sen} A.~Sen, ``$O(d) \otimes O(d)$ Symmetry of the Space of
Cosmological Solutions in String Theory, Scale Factor Duality,
and Two Dimensional Black Holes,'' Tata preprint, TIFR/TH/91-35, July 1991;
A.~Sen, ``Twisted Black $p$-brane Solutions in String
Theory,'' Tata preprint, TIFT/TH/91-37, August 1991;
S.~Hassan and A.~Sen,
``Twisting Classical Solutions in Heterotic String Theory,''
Tata preprint, TIFT/TH/91-40, September 1991;
G.~Veneziano, ``Scale Factor Duality for Classical and Quantum Strings,''
CERN preprint, CERN-TH-6077/91;
K.~Meissner and G.~Veneziano, ``Symmetries of Cosmological Superstring
Vacua,'' CERN preprint, CERN-TH-6138/91;
K.~Meissner and G.~Veneziano, ``Manifestly $O(d,d)$ Invariant Approach
to Space-Time Dependent String Vacua,'' CERN preprint, CERN-TH-6235/91,
September 1991.

\refis{Aichelburg} P.~Aichelburg and R.~Sexl,
\journal Gen. Rel. Grav., 2, 303, 1971.

\endreferences

\endit\end